# Analysis of Flame Acceleration Induced by Wall Friction in Open Tubes

V'yacheslav Akkerman[1,*], Chung K. Law[1], Vitaly Bychkov[2] and Lars-Erik Eriksson[3]

[1]*Department of Mechanical and Aerospace Engineering, Princeton University*
*Princeton, NJ 08544-5263, USA*

[2]*Department of Physics, Umeå University, 901 87 Umeå, Sweden*

[3]*Department of Applied Mechanics, Chalmers University of Technology*
*412 96 Gothenburg, Sweden*

**Abstract**

Spontaneous flame acceleration leading to explosion triggering in open tubes/channels due to wall friction was analytically and computationally studied. It was first demonstrated that the acceleration is effected when the thermal expansion across the flame exceeds a critical value depending on the combustion configuration. For the axisymmetric flame propagation in cylindrical tubes with both ends open, a theory of the initial (exponential) stage of flame acceleration in the quasi-isobaric limit was developed and substantiated by extensive numerical simulation of the hydrodynamics and combustion with an Arrhenius reaction. The dynamics of the flame shape, velocity, and acceleration rate, as well as the velocity profile ahead and behind the flame, have been determined.

**\* Corresponding author:**

V'yacheslav (Slava) Akkerman,

Dept. of Mechanical & Aerospace Engineering, Princeton University

Princeton, NJ 08544-5263 USA

Phone:  1 609 258 7643        Fax: 1 609 258 6233

**E-mail:** akkerman@princeton.edu

## 1. Introduction

The propagation speed ($S_L$) of laminar flames of hydrocarbon/air mixtures is much smaller than the sound speed ($c_S$) in the fresh gas, with $c_S / S_L = 10^3 \sim 10^4$ [1, 2]. As a result, the flow ahead of the flame is strongly subsonic and almost isobaric. However, even slow premixed flames may spontaneously accelerate, compress the unburned mixture and eventually trigger detonation. This phenomenon of deflagration to detonation transition (DDT) has been widely observed, particularly in experiments on flame propagation in tubes [3-7]. Indeed, the prevention or safe inducement of DDT is one of the key unsolved problems in combustion research.

A major cause of the flame acceleration is an increase in the flame surface area and thereby the total heat release rate. Various mechanisms can be envisioned in effecting such an increase: flame interaction with turbulent vortices [8-11], intrinsic flame instabilities [2, 12-15], ignition peculiarities [16, 17], flame-acoustic interactions [18, 19], etc. It is however noted that, for flames in tubes/channels, the role of turbulence and flame instabilities is rather supplementary as compared to either the Shelkin mechanism in smooth tubes [3, 7, 20-24] or acceleration in tubes with obstacles [25, 26]. As such, hereafter we shall focus on smooth tubes and on the Shelkin scenario of the DDT, i.e. on flame acceleration because of friction at the non-slip tube wall [3-7]. Mechanistically, thermal expansion of the burned gas generates a flow ahead of the flame, which becomes non-uniform because of friction at the wall. The nonuniform flow in turn renders the flame curved leading to its acceleration. The most important features of flame acceleration in channels, such as the exponential state of flame acceleration, the acceleration rate, and the self-similar flame shape, were obtained analytically, and substantiated by extensive direct numerical simulations [22, 23]. It was also shown that the flame dynamics depends strongly on the setup geometry. For example, while Bychkov et al. [22] obtained the flame acceleration from the *closed* end of a 2D channel, the simulation of Akkerman et al. [27] considered a 2D channel with both ends *open*, and demonstrated regular oscillations of a concave flame. Furthermore, a flame propagating from the *open* channel end to the *closed* one interacts with acoustic waves reflected form the closed end, leading to violent folding of the flame shape and, possibly, even flame turbulization [28, 29]. Thus we come to the question as whether a flame can accelerate only when propagating from the *closed* tube end to the *open* one, or whether we can obtain the flame acceleration and DDT in open tubes. This question is of considerable importance, for example, for safety issues. The question arises because while one should not expect flame acceleration in open tubes according to Ref. [27], this study was limited to a 2D planar geometry and it is recognized that 3D flows are more typical in reality. For example, quantitative investigation, both theoretical and computational, of the Shelkin scenario demonstrated that the flame acceleration from the closed end is much stronger in axisymmetric tubes in comparison with 2D channels, see Refs. [22, 23]. Therefore, one cannot rule out the possibility that flame oscillations in 2D open channels may be replaced by flame acceleration if we consider open tubes with 3D or axisymmetric flow geometry.

The main task of the present work is to study the flame dynamics in a *cylindrical tube* with *both ends open* and with non-slip adiabatic walls. Initially, we expected flame oscillations in this case, similar to regular flame pulsations in 2D channels [27]. In contrast, after some transitional time, we obtained the flame acceleration in an open tube which, however, is slower than that for a flame propagating from the closed end. We subsequently propose a criterion for flame acceleration, which involves the expansion factor in the burning process and a particular burning geometry. The criterion predicts the flame acceleration from the closed tube end both in 2D and axisymmetric cases, possibility of a steady front for 2D open channels (though the front may be unstable with respect to oscillations), and flame acceleration in axisymmetric open tubes. Thus, the criterion explains the previous simulation results of Refs. [22, 23, 27], as well as results of the present simulations. We then developed a theory of flame acceleration in axisymmetric open tubes, which predicts the acceleration rate, the shape of the accelerating flame, and the velocity profile ahead of it. The theory is substantiated by direct numerical simulations.

The paper consists of five technical sections. In Sec. 2 we estimate the criterion for flame acceleration in tubes/channel as a function of thermal expansion. The analytical theory of accelerating axisymmetric flames in cylindrical open tubes is developed in Sec. 3. In Sec. 4 we describe the direct numerical simulations, with the simulation results presented and discussed in Sec. 5. The details of the theory are presented in Appendices A and B.

## 2. Criterion for flame acceleration

As mentioned in the Introduction, direct numerical simulations demonstrated qualitatively different flame behaviors in open channels and tubes: flames oscillate in a 2D channel and accelerate in a cylindrical tube. To understand such a difference, in the present section we shall consider a hypothetical steady flame propagation in a tube/channel and determine the limits for the existence of such a steady state of burning. Figure 1 is a schematic of the study. A convex flame shape is described as

$$z_f(r,t) = z_{tip}(t) - f(r,t), \qquad (1)$$

where $z_{tip}(t)$ denotes the position of the flame tip, and the even function $f$ determines the deviation of the curved flame shape from the planar one. Motion of any local point of the flame surface is described as

$$dz_f / dt = U_{lab}(r,t) = U_u(r,t) + U_f(r,t), \qquad (2)$$

where $U_{lab}$ corresponds to the laboratory reference frame, $U_u$ the velocity field in the unburned gas, and $U_f$ the local flame velocity with respect to the unburned mixture. Averaging $U_f$ over the tube cross-section we obtain the total burning rate $U_w$. Assume that steady, or rather quasi-steady, flame propagation is possible. Then Eqs. (1) and (2) become

$$z_f(r) = U_{lab}t - f(r), \tag{3}$$

$$U_{lab}(r) = const = U_u(r) + U_f(r). \tag{4}$$

Thermal expansion across the flame produces a new gas volume $(\Theta-1)U_w A$ per unit time, where $\Theta \equiv \rho_u / \rho_b$ is the expansion factor, and $A$ the cross-section of the tube/channel. A flame pushes such a volume into the fuel mixture and the burnt gas. Thus the instantaneous average flows in the unburned and burnt gases may be expressed as

$$\langle U_u \rangle = \chi(\Theta-1)U_w, \qquad \langle U_b \rangle = -(1-\chi)(\Theta-1)U_w, \tag{5}$$

where $\langle ... \rangle$ stands for averaging over a fixed cross-section. Assuming Poiseuille flow ahead and behind the flame, we find the upstream $U_u$ and downstream $U_b$ velocity distributions as

$$U(r) = n\langle U \rangle \left(1 - \frac{r^2}{R^2}\right), \tag{6}$$

where $n = 3/2$ for a 2D channel and $n = 2$ for a cylindrical tube.

The factor $\chi$ in Eq. (5) depends on the setup configuration. If a flame propagates from the closed tube/channel end to the open one, then $\chi = 1$. In the opposite case of a flame propagating from the open tube end to the closed one, we have $\chi = 0$. In an open tube/channel, $\chi$ can be determined in the following manner. The momentum conservation equation takes the form [1, 2]

$$\frac{\partial}{\partial t} \iiint_V \rho \mathbf{U} dV = -\oiint_A \rho \mathbf{U}(\mathbf{U}d\mathbf{A}) + \mathbf{H}, \tag{7}$$

where the integrals are taken over the entire tube/channel volume and surface area, respectively, and $\mathbf{H}$ stands for the integral force acting on the fluid. In particular, $\mathbf{H} = \mathbf{H}_{iner} + \mathbf{H}_{fric}$, where $\mathbf{H}_{iner}$ designates the inertial force and $\mathbf{H}_{fric}$ is related to wall friction. Since there is no inertial force for steady flame propagation, we have $\mathbf{H}_{iner} = 0$. The friction force can be calculated by integrating the viscous stress over the entire tube/channel length

$$\mathrm{H}_{fric} = \int_L 2(\pi R)^{2n-3} \varsigma \left|\frac{\partial U}{\partial r}\right|_{r=R} dL = \frac{4n}{R}(\pi R)^{2n-3} \int_L \varsigma \langle U \rangle dL, \tag{8}$$

where $\varsigma$ is the dynamic viscosity. Assuming the viscosity coefficient to be constant, $\varsigma_u = \varsigma_b$, and that the flame is situated in the vicinity the tube center, at the distance $\approx L/2$ from both ends, we find

$$\mathrm{H}_{fric} = (2n/R)(\pi R)^{2n-3}(\Theta-1)(2\chi-1)\varsigma LU_w. \tag{9}$$

If the tube/channel is kept at rest, then the volume integral in the left-hand-side of Eq. (7) is zero; otherwise the flow would drag the tube/channel in the direction of the mean flow velocity due to viscous friction. The surface integral in Eq. (7) can be calculated as

$$\oiint_A \rho U(\mathbf{U} d\mathbf{A}) = 2R\left(\frac{\pi R}{2}\right)^{2n-3}\left[\chi^2 - \frac{(1-\chi)^2}{\Theta}\right](\Theta-1)^2 \rho_u U_w^2. \tag{10}$$

Finally, substituting Eqs. (9) and (10) into Eq. (7) we find

$$(\Theta-1)\left[\chi^2 - \frac{(1-\chi)^2}{\Theta}\right] = 2^{2n-3}\frac{\varphi}{\mathrm{Re}}\frac{S_L}{U_w}(2\chi-1), \tag{11}$$

with the aspect ratio $\varphi \equiv L/R$ and the flame propagation Reynolds number $\mathrm{Re} \equiv S_L R \rho_u / \varsigma$. The solution to Eq. (11) takes the form

$$\chi = C\left(\sqrt{1+C^{-1}}-1\right), \qquad C = \frac{1}{\Theta-1}\left(1 - \frac{2^{2n-3}}{\Theta-1}\frac{\varphi}{\mathrm{Re}}\frac{S_L}{U_w}\right). \tag{12}$$

When viscous effects dominate over the fluid motion, Eq. (12) yields $\chi \to 1/2$. In the opposite limit of long and narrow tubes/channels, $\varphi/\mathrm{Re} \ll 1$, Eq. (12) is reduced to

$$\chi = \left(1+\sqrt{\Theta}\right)^{-1}, \tag{13}$$

which yields $\chi = 0.26 \sim 0.31$ for typical hydrocarbon flames with $\Theta = 5 \sim 8$. Approximation (13) is adopted in this work.

In the limit of zero flame thickness, the local flame velocity is proportional to the local increase in the flame surface area

$$U_f(r) = S_L\sqrt{1+(df/dr)^2}. \tag{14}$$

Substituting Eqs. (5), (6), (14) into Eq. (4) we find

$$U_{lab} = S_L\sqrt{1+\left(\frac{df}{dr}\right)^2} + \chi(\Theta-1)nU_w\left(1-\frac{r^2}{R^2}\right). \tag{15}$$

Due to the locally planar flame tip, we have $df/dr = 0$ at the tube axis, where $r = 0$. Consequently, for $r = 0$, Eq. (15) is reduced to

$$U_{lab} = S_L + \chi(\Theta-1)nU_w. \tag{16}$$

On the other hand, averaging Eq. (15) along the front we find that a steady flame propagates in the laboratory reference frame as

$$U_{lab} = U_w + \chi(\Theta-1)U_w. \tag{17}$$

Equations (16) and (17) yield

$$U_w/S_L = \left[1-\chi(\Theta-1)(n-1)\right]^{-1}, \tag{18}$$

with $\chi$ determined by Eq. (12). According to Eq. (18), steady flame propagation is possible only for $\Theta < \Theta_c$, which implies that flames with larger thermal expansions have to accelerate because the right-hand-side of Eq. (18) becomes negative. In agreement with Eq. (13), the critical expansion factor $\Theta_c$ is given by

$$\Theta_c = \left(\frac{n}{n-1}\right)^2. \tag{19}$$

Since $\Theta_c = 9$ in a 2D channel, with $n = 3/2$, and $\Theta_c = 4$ in a cylindrical tube, with $n = 2$, and since typical thermal expansion is as large as $\Theta = 5 \sim 8$, the above result then indicates the strong possibility that a flame can accelerate in a cylindrical tube but will remain steady in a 2D channel because $\Theta > \Theta_c$ and $\Theta < \Theta_c$ in these two cases respectively.

It is emphasized that the result (19) is related to the approximation (13). In contrast, if the friction force dominates, $\chi = 1/2$, then the critical expansion factor is reduced to

$$\Theta_c = \frac{n+1}{n-1}, \tag{20}$$

which yields $\Theta_c = 5$ in a 2D channel and $\Theta_c = 3$ in a cylindrical tube. It is seen that accounting for the friction force does not noticeably change our prediction for $\Theta_c$ in an axisymmetric geometry.

Finally, if a flame propagates from a *closed* tube/channel end to the open one, then $\chi = 1$ and the criterion is replaced by

$$\Theta_c = \frac{n}{n-1}, \tag{21}$$

which yields $\Theta_c = 3$ for a 2D channel and $\Theta_c = 2$ for a cylindrical tube. Consequently, typical flames, with $\Theta = 5 \sim 8$, should accelerate in both tubes and channels if one end is closed. This result completely agrees with the theories and modeling [22, 23].

In addition, in light of the above results it is nevertheless noted that the critical expansion factors of Eqs. (19) and (20) formally correspond to an infinitely large $U_w$ in Eq. (18), while the realistic $\Theta_c$ could be much less than these estimations. Besides, the assumption of an infinitely thin flame could be quite restrictive in that the finite flame thickness would influence the flame dynamics. It is perhaps for these reasons that a steady flame obeying Eqs. (16) – (18) has not been observed in experiments or simulations, while direct numerical simulations of flames, with $\Theta = 8$, propagating in a 2D channel with open ends [27] demonstrated periodic oscillations of the flame. Consequently, a more definitive statement that we can make regarding the above results is that a flame with $\Theta > \Theta_c$ accelerates, while the behavior of flames with $\Theta < \Theta_c$ is not clear.

## 3. Theory of flame acceleration in open cylindrical tubes

In this section we shall derive analytically the primary characteristics of an accelerating flame in an axisymmetric geometry such as the state of acceleration, the acceleration rate, the flame shape and speed as well as the velocity profile in the flow generated by the flame propagation. We shall use the scaled variables $\eta = r/R$, $\xi = z/R$, $\mathbf{w} = \mathbf{u}/S_L$, $w_f = U_f/S_L$, $W_w = U_w/S_L$, and $\tau = t\,S_L/R$. The density and pressure are scaled by $\rho_u$ and $\rho_u S_L^2$, respectively. As in previous papers [22, 23], we assume that the flame-generated flow is plane-parallel. While this is an *approximation* because curvature of the flame front would induce transverse fluxes, it is justified by agreement between our previous and present theoretical predictions and numerical simulations. In particular, the parallel streamlines shown in Fig. 6 of Refs. [22, 23] and Fig. 4 of Ref. [27] demonstrate that the assumption of a plane-parallel upstream flow field is acceptable.

A plane-parallel flow, $w = \hat{\mathbf{e}}_z w_z(\eta, \tau)$, obeys the Navier-Stokes equation

$$\frac{\partial w}{\partial \tau} = \Pi(\tau) + \frac{1}{\mathrm{Re}} \frac{1}{\eta} \frac{\partial}{\partial \eta}\left(\eta \frac{\partial w}{\partial \eta}\right), \qquad (22)$$

where $\Pi = -\rho^{-1} \partial P / \partial \xi$. The assumption of a plane-parallel flow is self-consistent if the pressure gradient is a function of time only. The forcing $\Pi(\tau)$ is produced by the flame, which acts on the fresh gas as a piston moving with the velocity $W_w$ with respect to the fuel mixture. Due to the thermal expansion across the flame, the dimensionless gas volume increases by the value $\pi(\Theta-1)W_w$ per unit time. This volume is pushed partly into the fresh gas, and partly into the burnt gas; see Eq. (5). Although we considered steady burning in Sec. 2, the very concept of the factor $\chi$ separating flows in both directions also describes accelerating flames. If the state of flame acceleration is well-developed and self-similar, then the inertial forces can also be omitted, $\mathbf{H}_{iner}=0$. We note that if the factor $\chi$ varies with time depending on the flame position in the tube, the total burning rate $U_w$, and/or inertia, then the flame dynamics is not self-similar. Subsequently, in order to develop a self-similar analysis we take $\chi = const$. As a result, we have to make a choice between two limiting cases: whether hydrodynamic effect strongly dominates over viscous friction, Eq. (13), or vice versa. Recognizing that the relative role of the friction force decreases with increasing $U_w$, see Eq. (12), we shall choose the first option, $\chi = (1+\sqrt{\Theta})^{-1}$. Later, this choice will be substantiated by the simulation. Although the value $\chi$ is known we shall nevertheless keep the factor $\chi$ in all calculations. This allows us to generalize the theory of accelerating flames in cylindrical tubes. Indeed, by taking $\chi = 1$ we reproduce the situation of flame acceleration in tubes with one end closed. If $\chi$ is given by Eq. (13), or $\chi = 1/2$, then the theory below describes the flame acceleration in tubes with both ends open.

### 3.1. Flame-generated flow

The flow of the fuel mixture is nonuniform: the gas velocity vanishes at the non-slip wall because of friction and reaches its maximum value at the tube axis. Such a flow distorts the flame, which increases the total burning rate and, according to Eq. (5), leads to an additional increase in the flow velocity. The faster the flow, the stronger is the distortion of the flame shape, and the larger is the flame velocity. The flame accelerates as a consequence. In the limit of near-isobaric burning, the flame pushes weak compression waves, which have the same properties as acoustic waves. Then the pressure perturbations are proportional to the increase in the flame velocity, with $\Pi \propto W_w$. Assuming self-similar evolution of the flame shape and the flow, we can look for a solution to Eq. (22) in the form

$$w_u(\eta,\tau) = \alpha(\eta)\beta(\tau). \qquad (23)$$

Then Eq. (22) can be split into "space" and "time" equations

$$\sigma \operatorname{Re} \alpha = \frac{1}{\eta}\frac{d}{d\eta}\left(\eta \frac{d\alpha}{d\eta}\right) + \frac{\operatorname{Re}\Pi}{\beta}, \qquad \sigma = \frac{1}{\beta}\frac{d\beta}{d\tau} \qquad (24)$$

determining the velocity profile and the exponential acceleration as

$$\alpha = \alpha_{\max} \frac{I_0(\mu) - I_0(\mu\eta)}{I_0(\mu) - 1}, \qquad (25)$$

$$\beta = \beta_0 \exp(\sigma\tau), \qquad (26)$$

where $\mu \equiv \sqrt{\sigma \operatorname{Re}}$, and $I_0$ is the modified Bessel function of zero order. Obviously, $\alpha$ shows the shape of the flow, while $\beta$ describes its self-similar evolution. Consequently, Eqs. (5), (25) and (26) specify the plane-parallel velocity profiles ahead and behind the flame as

$$w_u(\eta,\tau) = \chi(\Theta-1)W_w \frac{I_0(\mu) - I_0(\mu\eta)}{I_0(\mu) - 2\mu^{-1}I_1(\mu)}, \qquad (27)$$

$$w_b(\eta,\tau) = -(1-\chi)(\Theta-1)W_w \frac{I_0(\mu) - I_0(\mu\eta)}{I_0(\mu) - 2\mu^{-1}I_1(\mu)} \qquad (28)$$

(see Appendix A for details), with the total flame velocity with respect to the fuel mixture $W_w(\tau) \propto \exp(\sigma\tau)$. The factor $\mu$ and the acceleration rate $\sigma$ are found next. In the laboratory reference frame, the total flame velocity is given by

$$W_{lab} = \langle w \rangle_u + W_w = [1 + \chi(\Theta-1)]W_w \propto \exp(\sigma\tau). \qquad (29)$$

Figure 2 presents the scaled velocity profile $\alpha/\alpha_{\max}$, Eq. (25), for various values of $\mu$. We observe that the velocity profile is very close to the Poiseuille flow for small $\mu$, $\mu = 0.1$.

Furthermore, even moderate values of $\mu$, $\mu = 3, 6$, resemble the same result. In contrast, the plot for $\mu = 15$ in Fig. 2 looks similar to an almost uniform flow with a transitional layer at the wall. We note that Eq. (25) formally coincides with that for flame propagation in a tube with a closed end. However, the values of $\mu$ calculated for the same $\Theta$ and Re in both configurations differ; see below. As a result, the absolute values of the flow velocities are strongly different in these configurations.

*3.2. Flame shape and velocity*

We now describe evolution of the convex flame in the above flow. Similar to Sec. 2, we look for the dimensionless function $\xi = \xi(\eta, \tau)$ describing the flame position in the form

$$\xi(\eta, \tau) = \zeta_{tip}(\tau) - F(\eta, \tau), \qquad (30)$$

where the function $\zeta_{tip}$ determines propagation of the flame tip, while the function $F$ describes the flame shape with respect to the tip, with $F(0, \tau) = 0$ by definition. Increase in the burning rate may be estimated by the growth of the flame surface. Every point on the flame propagates with respect to the fuel mixture in the $\xi$-direction with the local speed

$$w_f(\eta, \tau) = \sqrt{1 + (\partial F / \partial \eta)^2}, \qquad (31)$$

with $w_f(0, \tau) = 1$ and $\langle w_f \rangle = W_w$. In addition, the flame is convected by the flow, see Eq. (27). Consequently, in the laboratory reference frame the local shift of the flame surface is determined by the equation

$$d\xi / d\tau = w_u + w_f = w_u + \sqrt{1 + (\partial F / \partial \eta)^2}. \qquad (32)$$

The flame tip propagates as $d\zeta_{tip} / d\tau = w_u(0, \tau) + 1$. Substituting Eq. (30) into Eq. (32), we find the evolution equation for the flame shape

$$\partial F/\partial \tau = w_u(0,\tau) - w_u(\eta,\tau) + 1 - \sqrt{1+(\partial F/\partial \eta)^2}\,. \tag{33}$$

Due to the flame acceleration, after a short time we have $(\partial F/\partial \eta)^2 \gg 1$ everywhere except for the flat region close to the tube axis. Accounting for the fact that $\partial F/\partial \eta \geq 0$ for a convex flame, we can approximate Eq. (33) by the linear equation

$$\frac{\partial F}{\partial \tau} + \frac{\partial F}{\partial \eta} = w_u(0,\tau) - w_u(\eta,\tau). \tag{34}$$

Similar to the velocity profile, we look for solution to Eq. (34) in the form

$$F(\eta,\tau) = \Phi(\eta)\beta(\tau) \propto \Phi(\eta)\exp(\sigma\tau), \tag{35}$$

with $\Phi(0)=0$. Then Eq. (34) is reduced to

$$\sigma\Phi + \Phi' = 2\chi(\Theta-1)\left[\Phi(1) - \int_0^1 \Phi(\varepsilon)d\varepsilon\right] \frac{I_0(\mu\eta)-1}{I_0(\mu)-2\mu^{-1}I_1(\mu)}, \tag{36}$$

with the solution

$$\Phi(\eta) = \Phi_{max} \frac{[\sigma\Psi(\eta)+1]\exp(-\sigma\eta)-1}{[\sigma\Psi_{max}+1]\exp(-\sigma)-1}, \tag{37}$$

where $\Phi_{max} = \Phi(1)$ and

$$\Psi(\eta) = \int_0^\eta I_0(\mu\varepsilon)\exp(\sigma\varepsilon)d\varepsilon, \tag{38}$$

with $\Psi_{max} = \Psi(1)$. Integrating Eq. (37) in the domain $0<\eta<1$, we find an equation for the acceleration rate $\sigma$ and the factor $\mu = \sqrt{\sigma \mathrm{Re}}$ (see Appendix B for details)

$$\frac{I_0(\mu)-2\mu^{-1}I_1(\mu)}{2\chi(\Theta-1)} = \frac{(\sigma+1)\exp(-\sigma)-1}{\sigma^2} + \Psi_{max}\exp(-\sigma) - \int_0^1 \Psi(\eta)\exp(-\sigma\eta)d\eta. \tag{39}$$

Substituting $\chi=1$ into Eq. (39), we reproduce the result for flame propagation from the closed end of a cylindrical tube [23].

For arbitrary $\eta$, $\chi$ and $\mu$, Eq. (39) requires numerical solution. However, for realistically large thermal expansion $\Theta = 5 \sim 8$, the factor $\mu$ substantially exceeds unity, which allows decomposition in powers of $\mu^{-1}$. In the zero-order approximation of $\mu^{-1}$ we find

$$I_0(\mu) \approx \frac{\exp \mu}{\sqrt{2\pi\mu}}, \quad I_0(\mu\eta) \approx \frac{\exp(\mu\eta)}{\sqrt{2\pi\mu\eta}}, \quad \Psi_0(\eta) = \frac{\exp[(\mu_0 + \sigma_0)\eta]}{(\mu_0 + \sigma_0)\sqrt{2\pi\mu_0}}. \tag{40}$$

Then Eq. (39) is reduced to

$$\mu_0 + \sigma_0 = 2\chi(\Theta - 1), \tag{41}$$

with the solution

$$\mu_0 = \frac{\text{Re}}{2}\left(\sqrt{1 + \frac{8\chi(\Theta-1)}{\text{Re}}} - 1\right), \tag{42}$$

$$\sigma_0 = \frac{\mu_0^2}{\text{Re}} = \frac{\text{Re}}{4}\left(\sqrt{1 + \frac{8\chi(\Theta-1)}{\text{Re}}} - 1\right)^2. \tag{43}$$

In the limit of $\text{Re} \gg 8\chi(\Theta-1)$, Eqs. (42) and (43) are further simplified to

$$\mu_0 = 2\chi(\Theta - 1), \quad \sigma_0 = 4\chi^2(\Theta-1)^2/\text{Re} \ll \mu_0. \tag{44}$$

In particular, for $\chi$ determined by Eq. (13) and typical alkane hydrocarbon flames with $\Theta \approx 8$, we have $\mu_0 = 2(\sqrt{\Theta} - 1) \approx 3.7$ for $\text{Re} \gg 1$. Consequently, one should expect the accuracy of the zero-order approximation of $\mu^{-1}$ to be about 30%. We note that the result is more accurate for a tube with a closed end. Indeed, taking $\chi = 1$ we find $\mu_0 = 14$, which corresponds to an accuracy of 7%.

Figure 3 shows the scaled flame shape $\Phi(\eta)/\Phi_{\max}$ determined by Eq. (37) for the expansion factor $\Theta = 8$ and various flame propagation Reynolds numbers, $\text{Re} = 10, 50, 200$. All three plots demonstrate qualitatively the same convex "U"- shaped flames. The larger Re, the wider is the flame tip and the narrower is the transitional region at the wall. Still, the dependence of the flame shape versus the Reynolds number is not so strong at high $\text{Re}$: the plots for $\text{Re} = 50$ and $\text{Re} = 200$ are not clearly distinguished.

## 4. Direct numerical simulations

To validate the theoretical predictions of Sec. 3, we performed extensive numerical simulation of the hydrodynamics and combustion with an Arrhenius reaction. Similar to the theory, an axisymmetric flame propagating in a cylindrical tube with both ends open and with non-slip adiabatic wall was considered. The basic equations, the description of the solver and the numerical method are presented, in particular, in Refs. [23, 31-33].

The fuel mixture is assumed to be a perfect gas of constant heat capacities, $C_V = 5R_p / 3m$, $C_P = 7R_p / 3m$, where $m = 2.9 \times 10^{-2}$ kg/mol is the molecular weight and $R_P \approx 8.31$ J/(mol·K) the perfect gas constant. The equation of state is $P = \rho R_p T / m$. We chose the initial pressure and temperature of the fuel as $P_u = 10^5$ Pa and $T_u = 300$ K. Chemical kinetics was approximated by an irreversible one-step Arrhenius reaction of the first order, with an activation energy $E_a$, a characteristic collision time constant $\tau_R$ and the energy release in the reaction $Q$. We chose $E_a$, $Q$, $\tau_R$ in such a manner as to obtain the planar flame speed $S_L = 34.7$ cm/s. Then a typical flow velocity is much smaller than the sound speed, $c_S$, with the flame propagation Mach number $M = S_L / c_S = 10^{-3}$. This corresponds to a strongly subsonic flow, which may be described as isobaric, with the thermal expansion coupled to the energy release in the burning process as

$$\Theta = T_b / T_u = 1 + Q / C_P T_u. \tag{45}$$

We chose the activation energy $E_a = 7R_p T_b$, which allows smoothing the reaction zone over several computational cells. In most simulation runs, thermal expansion was chosen as $\Theta = 8$, which is typical for the burning of alkane hydrocarbons. However, other expansion factors $\Theta = 2 \sim 5$ were also considered. We chose the dynamic viscosity $\varsigma = 2.38 \times 10^{-5}\,\text{Ns/m}^2$. To avoid the diffusional-thermal instability we assumed the coefficients of thermal diffusivity and fuel diffusion to be equal, i.e. unity Lewis number, $Le \equiv Sc/Pr = 1$, with the Prandtl and Schmidt numbers being $Sc = Pr = 0.7$. The instantaneous total burning rate was calculated as [30]

$$U_w = \frac{1}{\rho_u \pi R^2} \int \frac{\rho Y}{\tau_R} \exp\left(-E_a / R_P T\right) 2\pi r\, dr\, dz, \qquad (46)$$

where $Y$ is the local mass fraction of the fuel mixture. The integral in Eq. (46) is taken over the entire tube. It was demonstrated that the value $U_w / S_L$ calculated with Eq. (46) correlates very well with the scaled total flame velocity with respect to the fuel mixture and the scaled surface area of the flame isotherms [11, 30].

The characteristic width of the burning zone may be estimated as $L_f = \varsigma / \Pr \rho_f S_L$. Then the ratio of the hydrodynamic and chemical length scales is determined by the Peclet number $Pe = R / L_f$, which is related to the flame propagation Reynolds number as $Pe = RePr$, i.e. $Pe = 0.7Re$ in the present simulation. Consequently, the theoretical predictions of Fig. 3 are related to $Pe = 7, 35, 140$. The parameter $L_f$ also determines the size of the calculation grid. As a result, the computational time is proportional to $Pe^3$.

We took the tube length much larger than the tube radius, $L = 400R = (2 - 14) \times 10^3 L_f$, so as to simulate an infinitely long tube. We emphasize that the simulation results do not depend on the tube length. The grid was rectangular, with the grid wall parallel to the radial and axial directions. It was uniform along the radial direction, with the cell size equal to $0.5L_f$. To perform all the calculations in a reasonable time, we made the grid nonuniform along the $z$ - axis, with the zone of fine mesh around the flame. In that region, the grid size was $0.2L_f$ in the $z$ - direction, which resolves quite well the internal flame structure. The length of the fine grid zone $L_g$ must be large enough to contain the flame during the entire simulation run. The region of the fine grid is chosen in such a manner that the initial position of the flame is situated inside this zone, at the distance $50L_f$ from its left border. Outside the region of the fine grid, the mesh size grows gradually with $\approx 3\%$ change in size between the neighboring cells. Using a gradually growing grid, we managed to take ultimately long tubes, so the fine mesh zone, namely the distance that the flame propagates for the whole simulation run, was much smaller than the entire tube length. We tested numerical resolution widely in our previous studies; see, for instance, Fig. 1 in Ref. [27]. Furthermore, in the present work, we have performed the test simulation run with the square mesh $0.2L_f \times 0.2L_f$ in the fine grid zone $L_g$. The difference between the results of the two simulations is quite small, which justifies that our standard resolution $0.5L_f \times 0.2L_f$ is appropriate. Initial states of the unburned inflow and burned outflow are uniform, respectively given by

$$\rho = \rho_u, \quad T = T_u, \quad u_z = -S_L, \quad u_r = 0, \quad Y = 1, \tag{47}$$

$$\rho = \rho_u / \Theta, \quad T = \Theta T_u, \quad u_z = -\Theta S_L, \quad u_r = 0, \quad Y = 0. \tag{48}$$

To avoid the influence of sound waves and weak shocks reflected from the open tube ends we applied the non-reflecting boundary conditions at the ends. As the initial conditions, we used the Zeldovich-Frank-Kamenetski solution for a planar flame [5]. An initially planar flame was created at the distance about $200R$ from the left tube end. Finally, we adopted adiabatic ($\hat{\mathbf{n}} \cdot \nabla T = 0$) and non-slip ($\mathbf{u} = 0$) boundary conditions at the tube wall.

## 5. Results and discussion

In this section we shall present and discuss results of the simulation, and compare them to the theoretical prediction of Sec. 3. We simulated flows in tubes of $Pe = R/L_f = 5 \sim 35$, which corresponds to the flame propagation Reynolds number $\text{Re} \approx 7 \sim 50$. It is noted that the Reynolds number related to the *flow* could be 1–2 orders of magnitude larger,

$$\text{Re}_{flow,u} \approx \chi(\Theta-1)(U_w/S_L)\text{Re} \quad \text{or} \quad \text{Re}_{flow,b} \approx (1-\chi)(\Theta-1)(U_w/S_L)\text{Re}, \quad (49)$$

which are nevertheless still below the value for the transition to turbulence. In the simulations, we started with an initially planar flame shape, which subsequently becomes distorted in a short time due to interaction with the wall and the flow.

The characteristic behavior of the shape of the flame with $\Theta = 8$ is presented in Fig. 4a for a tube of radius $R = 10L_f$ and in Fig. 4b for a relatively wider tube with $R = 30L_f$. Figure 4a shows that the flame acquires a strongly curved, concave shape by instant (III), which is accompanied by noticeable flame acceleration. The acceleration subsequently stops and is followed by deceleration. As a result, the flame at instant (IV) becomes much flatter, with the total burning rate slightly exceeding $S_L$. The flame shape then inverts from concave to convex, at instant (V). This convex flame starts accelerating in a self-similar manner without bound, as shown at instants (VI) – (VIII). A similar tendency is also shown in Fig. 4b for the tube of radius $R = 30L_f$. Since the flame in Fig. 4b is much thinner than the tube radius, the individual isotherms are not clearly distinguished. In contrast, the size of the preheat zone is of the same order of magnitude as the tube radius in Fig. 4a, and hence the isotherms are distinguished. Furthermore, the flame is curved much stronger, and variations of the flame shape are much more pronounced in the wider tube, where the convex flame acquires a noticeable cusp at the axis. The same effect has been observed in a tube with a closed end. Such a cusp is presumably related to the development of the Darrieus-Landau and/or Rayleigh-Taylor instabilities at a planar flame tip. The contribution of such a cusp into the total flame surface area and the flame speed is however quite small, being $\propto (r_c/R)^2$, where $r_c$ is the cusp radius. At the stage of self-similar flame acceleration, the flame isotherms of Fig. 4 resemble the theoretical prediction (35) shown in Fig. 3, except for the cusp at the axis as we do not account for flame instabilities in the theory.

For comparison, in Fig. 5 we present the flame dynamics in a 2D open channel of half-width $20L_f$ simulated in Ref. [27]. It is seen that, at the initial stage of burning, the flame dynamics in Fig. 5 resembles that of Fig. 4, with the acceleration-deceleration of a concave flame in positions (a) – (d) in Fig. 5. The flame shape in positions (c) and (d) in Fig. 5 resembles that in position (V) in Fig. 4a. However, the concave-to-convex inversion does not occur in the 2D channel. In contrast, the flame remains concave and accelerates again after some time, as shown in position (e) in Fig. 5, which is followed by one more flame deceleration; see position (f). The flame shapes in positions (c) and (f) in Fig. 5 appear identical. Consequently, we have regular oscillations of a concave flame in a channel instead of flame acceleration in a tube. We have explained this effect in Sec. 2, where the criterion for flame acceleration versus "quasi-steady" flame propagation is developed. However, the analysis of Sec. 2 is correct for steady flame propagation, while in the case of unsteady propagation the inertia of the upstream and downstream gases influences the flame dynamics. It is then reasonable to anticipate that inertial effects could be responsible for the initial flame behavior in the simulations of Figs. 4 a, b, and for the entire set of oscillations in Fig. 5. Subsequently, we observe a "competition" between inertial force leading to flame oscillations and the viscous force resulting in the flame acceleration. The inertial effects are important at the initial stage, while the viscous force dominates for self-similar flame acceleration at later stages. We also anticipate that the initial flame behavior results similarly from a "competition" between the viscous force and the hydrodynamic effects.

The last stage of the flame evolution in Figs. 4 a, b is qualitatively similar to the flame dynamics in a tube or channel with $\chi = 1$. In Figure 6 we show flame acceleration in a cylindrical tube of radius $R = 25L_f$ with a closed end simulated in Ref. [23]. Indeed, the flame acceleration in Fig. 6 resembles qualitatively Fig. 4b, including a cusp at the tube axis. Nevertheless, the acceleration is much weaker in an open tube, in which the flame-generated flow is distributed between the fresh and burnt gas. If the "burnt" end of the tube is closed, then the entire flame-generated gas volume is pushed towards the fresh gas, which leads to a much larger acceleration rate than that in a tube with both ends open. Indeed, Fig. 4a shows that, in an open tube of $Pe = 10$, the flame propagates through $\approx 8R$, and its surface area increases $\approx 4$ times during $t \approx 1.4R/S_L$. When the tube radius is $R = 30L_f$, Fig. 4b, the flame propagates through $\approx 13R$, and the flame surface area increases $\approx 3$ times during $t \approx 1.8R/S_L$. In contrast, according to Fig. 6, the flame in the tube of $Pe = 25$ propagates through $\approx 20R$ and accelerates $\approx 10$ times during a much shorter time interval, $t \approx 0.7R/S_L$, if $\chi = 1$.

The complete evolution of the flame in a cylindrical open tube of radius $R = 30L_f$ is shown in snapshots in Fig. 7. Again, we observe acceleration-deceleration of a concave flame, the concave to convex inversion and acceleration of a convex flame. Figure 8 presents evolution of the scaled flame velocity $U_w/S_L$ for $\Theta = 8$ and $R/L_f = 5 \sim 35$. The flame velocity is given by Eq. (46). Figures 8 (a – d) correspond to the three main stages of the flame dynamics. Initial acceleration-deceleration of concave flames is shown in Fig. 8a, demonstrating that the wider the tube, the stronger is the flame corrugation and the burning rate. This is basically in line with Fig. 5. Figure 8b is related to an "intermediate" stage around the concave to convex flame inversion, while Fig. 8c describes the stage of self-similar convex flame acceleration. An opposite tendency is shown here in that the wider the tube, the weaker is the flame acceleration. Figure 8d is the counterpart of Fig. 8c plotted in the semi-logarithmic scale, which is used in to demonstrate the exponential state of flame acceleration and to measure the acceleration rate $\sigma$ (the angular coefficients of the lines in Fig. 8d). Two dotted plots in Figs. 8 c, d are related to tubes of $Pe = 5, 35$ with a closed end. We observe again that the acceleration is much stronger if $\chi = 1$. The dashed plot in Fig. 8 (a – d) is related to the test simulation run for $Pe = R/L_f = 10$ with the square mesh $0.2L_f \times 0.2L_f$ in the fine grid zone. The difference between the results of the two simulations is quite small, which justifies that our standard resolution $0.5L_f \times 0.2L_f$ is appropriate.

It is noted that the dynamics of the flame velocity in Fig. 8 correlates well with the evolution of the flame shape in Figs. 4 and 7. In particular, for $Pe = 10$, the scaled flame velocity achieves a local (concave) maximum $U_w \approx 1.2 S_L$ at $t \approx 0.4 R/S_L$; it decreases to almost $S_L$ at $t \approx 0.8 R/S_L$, and then the convex flame accelerates, achieving $U_w \approx 1.2 S_L$ again at $t \approx R/S_L$. These three instants are related to the positions (III), (V) and (VI) in Fig. 4a. The surface area of the flame profiles in Figs. 4b, 7 also correlates with the plots for $Pe = 30$ in Figs. 8 (a – d). Furthermore, when the flame is strongly elongated, $z_{max} - z_{min} >> R$, then the scaled flame speed may be estimated as

$$U_w / S_L \approx (z_{max} - z_{min})/R. \qquad (50)$$

Here $z_{min}$ and $z_{max}$ are the left and right borders of the flame (at the wall, and close to the tube axis, respectively), which means that $Y \approx 0$, $T \approx 2400\,\text{K}$ at $z < z_{min}$ and $Y \approx 1$, $T \approx 300\,\text{K}$ at $z > z_{max}$ for any $r$. Anticipating that the inertia of the upstream/downstream gas is responsible for the initial flame behavior (Fig. 8a), it can then be suggested that the inertial effects may be neglected compared to viscous forces when the flame starts to propagate in a self-similar manner, i.e. for $U_w / S_L = 3 - 5$ at $\Theta = 8$.

The factor $\sigma$, calculated from Fig. 8d, is shown by symbols in Fig. 9 as a function of the Peclet number. The theoretical prediction of Sec. 3, with $\chi$ given by Eq. (13), is also shown in Fig. 9 with the solid line describing the numerical solution to Eq. (39) and the dashed line describing the zero-order approximation in $\mu^{-1}$, see Eq. (43). The symbols agree well with the theoretical prediction. Nevertheless, the agreement is less accurate than that in previous studies [22, 23], implying that the accuracy of the theory developed in Sec. 3 is determined by the factor $\mu$: the larger the $\mu$, the stronger is the condition $\mu^{-1} \ll 1$ and the approach of self-similar flame dynamics. Indeed: we have $\mu \approx 3 \sim 5$ in the present work. Previous studies of flame acceleration in a 2D channel with a closed end [22] used $\mu \approx 6 - 8$ and demonstrated much better agreement between theory and simulations. Finally, theory and modeling of flames in a cylindrical tube with a closed end [23] used $\mu \approx 12 - 15$ and demonstrated even better agreement than Ref. [22].

We emphasize that, unlike the velocity profiles in Fig. 8, the parameter $\sigma$ does not depend on the initial conditions because the flame behavior is scale invariant. According to the theory of Sec. 3, the acceleration rate is determined mainly by the scaled tube width and the thermal expansion in the burning process; $\sigma$ increases with $\Theta$ and decreases with $Pe$. In very wide tubes, $R \gg L_f$, the acceleration rate tends to the asymptotic value $\sigma = 4(\sqrt{\Theta} - 1)^2 / \text{Re}$, see Eq. (44)

In the present simulation, we also studied the effect of thermal expansion on the flame dynamics. Evolution of the scaled flame velocity $U_w / S_L$ for various expansion factors $\Theta = 2 \sim 8$ is shown in Fig. 10 for a tube of radius $R = 10L_f$ and in Fig. 11 for a relatively wider tube with $R = 20L_f$. As in Fig. 8, Fig. 11a presents the initial acceleration-deceleration of concave flames, while the later (convex) stage of the flame dynamics is presented in Fig. 11b. We observe that the larger thermal expansion, the stronger is the flame corrugation and the burning rate for both concave and convex flames. Furthermore, the larger $\Theta$, the faster the concave-to-convex transition occurs. Figures 10, 11b demonstrate quite weak, almost linear flame acceleration for small $\Theta$ instead of the exponential increase observed for $\Theta = 5 \sim 8$, and the threshold between exponential and linear acceleration states is in between $\Theta = 3$ and $\Theta = 5$, which agrees with our prediction for $\Theta_c = 3 \sim 4$, Eqs. (19), (20). We emphasize however that steady flame propagation discussed in Sec. 2 has not been observed in the simulation, although the ultimately weak acceleration for $\Theta = 2$ in Figs. 10, 11b can be interpreted in a certain sense as slow saturation to the steady flame propagation. Among the possible reasons for such a discrepancy, the viscous friction force can be of primary importance at the initial stage, so the model (13) does not work. However, the relative role of friction decreases with the flame acceleration, Eq. (12). Then, at a certain stage, the role of the friction force becomes minor, so the self-similar exponential acceleration predicted by the theory is also observed in the simulation.

Consequently, the only definitive statement that we can make regarding the above results is that a flame with $\Theta > \Theta_c$ accelerates in a self-similar manner, which is supported by the model of Sec. 2, the theory of Sec. 3, and the computational results of Sections 4 – 5.

## 6. Summary

The dynamics of flame evolution in cylindrical tubes with non-slip adiabatic wall was studied analytically and computationally. We compared the flame dynamics in 2D channels and tubes; as well as in channels/tubes with one end and both ends open. We demonstrated that convex flames tend to accelerate in both tubes and channels. An initially planar flame acquires a concave shape and oscillates regularly in a 2D open channel. In contrast, a concave flame in an open tube inverts into a convex front after the first oscillation, which is followed by exponential flame acceleration in a self-similar manner. In Sec. 2 an explanation was advanced for such a qualitative difference in the flame behavior. We also determined the criterion for the flame acceleration depending on the combustion configuration and the expansion factor; see Eqs. (19), (20). The flame accelerates if $\Theta > \Theta_c$, but may oscillate if $\Theta < \Theta_c$, and the critical expansion factors $\Theta_c$ for tubes and channels differ strongly. In Sec. 3, we developed an analytical theory of accelerating axisymmetric flames in cylindrical open tubes, extending the previous analysis of Akkerman et al. [23] but accounting for the fact that the flame propagation generates a flow in both directions if the tube ends are open. We determined the main properties of flame acceleration such as the upstream and downstream velocity profiles, Eqs. (27), (28), Fig. 2, the flame evolution, Eqs. (30), (32), and shape, Eqs. (35), (37), (38), Figs. 3, and the acceleration rate $\sigma$ as a solution to Eq. (39), Fig. 9. The theory was validated by extensive numerical simulations of the hydrodynamics and combustion with an Arrhenius reaction. The simulation results agree with the theory; see Fig. 9. Since the flame accelerates due to the thermal expansion in the process of burning, the acceleration rate should increase with the expansion factor, which agrees with the analytical estimation (43) and is proved by numerical simulations; see Figs. 10, 11. On the other hand, $\sigma$ decreases with the Reynolds/Peclet number of the flow. Inspecting Figs. 4 (a, b) and 7, we conclude that the flame shape remains self-similar and does not depend on the initial conditions as soon as the exponential state of flame acceleration (Fig. 8 c, d) is achieved.

It is noted that in the present study the flame accelerates exponentially while the flow is almost isobaric. As the acceleration rate $\sigma$ for flames in open tubes obtained in the present work is moderate, the flame acceleration remains exponential during almost all simulation runs with $\Theta = 8$. The exponential growth rate starts to be saturated at the end of the simulation run only for fast acceleration in a tube with $R = 5L_f$. Consequently, the isobaric approach in the simulations is valid, allowing comparison of the simulation results to the theoretical predictions of Sec. 3. However, as soon as the flow velocity becomes comparable to the sound speed, influence from the flame-generated compression waves becomes important, thereby reduces the flame acceleration [7, 34, 35].

**Acknowledgments**

This work was supported by the U.S. Air Force Office of Scientific Research and by the Swedish Kempe Foundation.

**Appendix A. Upstream/downstream velocity profiles**

Here we demonstrate how the formulas for the flame-generated velocity profiles, (25) – (28), are derived. Substituting Eq. (23) into Eq. (22) we find

$$\alpha(\eta)\frac{1}{\beta}\frac{d\beta}{d\tau} = \frac{\Pi(\tau)}{\beta(\tau)} + \frac{1}{\text{Re}}\frac{1}{\eta}\frac{d}{d\eta}\left(\eta\frac{d\alpha}{d\eta}\right), \tag{A1}$$

which can be rewritten in the form of two equations (see Eq. (24))

$$\sigma = \frac{1}{\beta}\frac{d\beta}{d\tau}, \tag{A2}$$

$$\sigma \text{Re}\,\alpha = \frac{1}{\eta}\frac{d}{d\eta}\left(\eta\frac{d\alpha}{d\eta}\right) + C_\Pi, \tag{A3}$$

with $\beta(0) = \beta_0$, $\alpha(1) = 0$, $\alpha(0) = \alpha_{max}$ and $C_\Pi = \text{Re}\,\Pi/\beta$ (recall that $\Pi \propto W_w \propto \beta$, so $C_\Pi$ is a constant). The solution to Eq. (A2) is given by

$$\beta = \beta_0 \exp(\sigma\tau). \tag{A4}$$

To solve Eq. (A3), we introduce $\mu = \sqrt{\sigma \text{Re}}$, $\tilde{\eta} = \mu\eta$ and $\tilde{\alpha} = \alpha - C_\Pi/\mu^2$. Then Eq. (A3) takes the form

$$\tilde{\alpha} = \frac{1}{\tilde{\eta}}\frac{d}{d\tilde{\eta}}\left(\tilde{\eta}\frac{d\tilde{\alpha}}{d\tilde{\eta}}\right). \tag{A5}$$

By definition, the modified Bessel function of zero order $I_0(\eta)$ is a solution to Eq. (A5), with $I_0(0) = 1$. Consequently, $\tilde{\alpha} = C_\alpha I_0(\tilde{\eta})$, i.e. $\alpha = C_\alpha I_0(\mu\eta) + C_\Pi/\mu^2$. Accounting for the conditions at $\eta = 0$ and $\eta = 1$, we find $C_\alpha = -[1 - I_0(\mu)]^{-1}$ and $C_\Pi = I_0(\mu)[1 - I_0(\mu)]^{-1}$, and finally obtain Eq. (25)

$$\alpha/\alpha_{max} = \frac{I_0(\mu) - I_0(\mu\eta)}{I_0(\mu) - 1}. \tag{A6}$$

Equation (A6) determines the scaled velocity profile of the plane-parallel flows both upstream and downstream the flame. In the limit of $\mu\eta >> 1$, it is reduced to

$$\alpha/\alpha_{max} \approx 1 - \eta^{-1/2} \exp[\mu(\eta-1)]. \tag{A7}$$

Equation (A7) describes the plot $\mu = 15$ in Fig. 2 quite well. In the opposite case of $\mu \to 0$, Eq. (A6) reproduces the parabolic, Poiseuille flow.

Equations (A4) and (A6) for either upstream or downstream flow yield

$$w = \alpha_{max} \beta_0 \exp(\sigma\tau) \frac{I_0(\mu) - I_0(\mu\eta)}{I_0(\mu) - 1}. \tag{A8}$$

Averaging Eq. (A8) along the flame front we find

$$\langle w \rangle = \int_0^1 w \, 2\eta \, d\eta = \frac{\alpha_{max} \beta_0 \exp(\sigma\tau)}{I_0(\mu) - 1} \left[ I_0(\mu) - \int_0^1 I_0(\mu\eta) 2\eta \, d\eta \right]. \tag{A9}$$

Using the definition of the modified Bessel function, Eq. (A5), and the relation $I_1(\eta) = dI_0(\eta)/d\eta$, we calculate the integral in Eq. (A9) as

$$\int_0^1 I_0(\mu\eta) 2\eta \, d\eta = \frac{1}{\mu^2} \int_0^1 \frac{1}{\eta} \frac{\partial}{\partial \eta}\left( \eta \frac{\partial I_0(\mu\eta)}{\partial \eta} \right) 2\eta \, d\eta = \frac{2\eta}{\mu^2} \frac{\partial I_0(\mu\eta)}{\partial \eta}\bigg|_0^1 = \frac{2}{\mu} I_1(\mu). \tag{A10}$$

Then Eq. (A9) takes the form

$$\langle w \rangle = \alpha_{max} \beta_0 \exp(\sigma\tau) \frac{I_0(\mu) - 2\mu^{-1} I_1(\mu)}{I_0(\mu) - 1}. \tag{A11}$$

Division of Eq. (A8) by Eq. (A11) yields

$$w = \langle w \rangle \frac{I_0(\mu) - I_0(\mu\eta)}{I_0(\mu) - 2\mu^{-1} I_1(\mu)}. \tag{A12}$$

Then substituting Eq. (A12) into Eq. (5) we find the instantaneous velocity distribution upstream and downstream of the flame (see Eqs. (27), (28))

$$w_u(\eta,\tau) = \chi(\Theta - 1) W_w \frac{I_0(\mu) - I_0(\mu\eta)}{I_0(\mu) - 2\mu^{-1} I_1(\mu)}, \tag{A13}$$

$$w_b(\eta,\tau) = -(1-\chi)(\Theta-1)W_w \frac{I_0(\mu) - I_0(\mu\eta)}{I_0(\mu) - 2\mu^{-1}I_1(\mu)}. \tag{A14}$$

Maximal/minimal values are achieved at the tube axis, where $\eta = 0$.

## Appendix B. Flame shape and velocity

Here we derive equations (37) – (39) for the flame shape and the factor $\mu$ (or $\sigma$). Flame propagation is described by Eq. (30), $\xi = \xi_{tip} - F(\eta,\tau)$, and evolution of the flame shape is approximated by Eq. (34),

$$\partial F/\partial \tau + \partial F/\partial \eta = w_a(0,\tau) - w_a(\eta,\tau). \tag{B1}$$

We look for the solution to Eq. (34) in the form (35),

$$F(\eta,\tau) = \Phi(\eta)\exp(\sigma\tau), \tag{B2}$$

with $\Phi(0) = 0$. Substituting Eqs. (A13), (B2) into Eq. (B1) we find

$$(\sigma\Phi + \Phi')\exp(\sigma\tau) = \chi(\Theta-1)W_w \frac{I_0(\mu\eta) - 1}{I_0(\mu) - 2\mu^{-1}I_1(\mu)}. \tag{B3}$$

According to Eq. (31), the instantaneous burning rate in Eq. (B3) can be calculated as

$$W_w = \left[\int_0^1 \sqrt{1 + \left(\frac{\partial F}{\partial \varepsilon}\right)^2}\, 2\pi\varepsilon\, d\varepsilon\right] \times \left[\int_0^1 2\pi\varepsilon\, d\varepsilon\right]^{-1} \approx$$

$$2\int_0^1 \frac{\partial F(\varepsilon,\tau)}{\partial \eta} \varepsilon\, d\varepsilon = 2\left[\Phi(1) - \int_0^1 \Phi(\varepsilon)d\varepsilon\right]\exp(\sigma\tau). \tag{B4}$$

Then Eq. (B3) takes the form

$$\sigma\Phi + \Phi' = 2\chi(\Theta-1)\left[\Phi(1) - \int_0^1 \Phi(\varepsilon)d\varepsilon\right]\frac{I_0(\mu\eta) - 1}{I_0(\mu) - 2\mu^{-1}I_1(\mu)}, \tag{B5}$$

with the solution

$$\Phi(\eta) = \tilde{C}[(\sigma\Psi(\eta) + 1)\exp(-\sigma\eta) - 1], \tag{B6}$$

where

$$\Psi(\eta) = \int_0^\eta I_0(\mu\varepsilon)\exp(\sigma\varepsilon)d\varepsilon, \qquad (B7)$$

$\Psi(0) = 0$, and the numerical factor $\tilde{C}$ obeys the equation

$$\tilde{C} = \frac{2}{\sigma}\frac{\chi(\Theta-1)}{I_0(\mu)-2\mu^{-1}I_1(\mu)}\left[\Phi_{max} - \int_0^1 \Phi(\varepsilon)d\varepsilon\right]. \qquad (B8)$$

Obviously, Eq. (B6) may be also rewritten in the scaled form (Eq. (37))

$$\Phi(\eta) = \Phi_{max}\frac{[\sigma\Psi(\eta)+1]\exp(-\sigma\eta)-1}{[\sigma\Psi_{max}+1]\exp(-\sigma)-1}. \qquad (B9)$$

Here $\Phi_{max} = \Phi(1)$ and $\Psi_{max} = \Psi(1)$ are the maximum values of $\Phi$ and $\Psi$ achieved at the tube wall. Integrating Eq. (B6) in the domain $0 < \eta < 1$ we find

$$\int_0^1 \Phi(\eta) = \tilde{C}\left[\sigma\int_0^1 \Psi(\eta)\exp(-\sigma\eta)d\eta + \int_0^1 \exp(-\sigma\eta)d\eta - 1\right]. \qquad (B10)$$

At the tube wall, $\eta = 1$, Eq. (B6) yields

$$\Phi_{max} = \tilde{C}[(\sigma\Psi_{max}+1)\exp(-\sigma)-1]. \qquad (B11)$$

Subtracting Eq. (B10) from Eq. (B11) and using Eq. (B8) we obtain

$$\left[\Phi_{max} - \int_0^1 \frac{2\chi(\Theta-1)/\sigma}{I_0(\mu)-2\mu^{-1}I_1(\mu)}\left\{\Phi_{max} - \int_0^1 \Phi(\eta)d\eta\right\} \times \Phi(\eta)d\eta\right] =$$

$$\left[\sigma\left(\Psi_{max}\exp(-\sigma) - \int_0^1 \Psi(\eta)\exp(-\sigma\eta)d\eta\right) + \exp(-\sigma) - \int_0^1 \exp(-\sigma\eta)d\eta\right], \qquad (B12)$$

which is transformed into Eq. (39)

$$\frac{I_0(\mu)-2\mu^{-1}I_1(\mu)}{2\chi(\Theta-1)} = \frac{(\sigma+1)\exp(-\sigma)-1}{\sigma^2} + \Psi_{max}\exp(-\sigma) - \int_0^1 \Psi(\eta)\exp(-\sigma\eta)d\eta. \qquad (B13)$$

**References**

[1]  F. A. Williams, *Combustion Theory* (Benjamin, Redwood City, CA, 1985).

[2]  C. K. Law, *Combustion Physics* (Cambridge University Press, New York, NY, 2006).

[3]  K. I. Shelkin, "Influence of tube walls on detonation ignition," Zh. Eksp. Teor. Fiz. **10**, 823 (1940).

[4]  P. A. Urtiew and A. K. Oppenheim, "Experimental observations of the transition to detonation in explosive mixtures," Proc. R. Soc. Lond. A **295**, 13 (1966).

[5]  Ya. B. Zeldovich, G. I. Barenblatt, V. B. Librovich, and G. M. Makhviladze, *Mathematical Theory of Combustion and Explosion* (Consultants Bureau, New York, NY, 1985).

[6]  G. D. Roy, S. M. Frolov, A. A. Borisov, and D. W. Netzer, "Pulse detonation propulsion: challenges, current status, and future perspective," Prog. En. Combust. Sci. **30**, 545 (2004).

[7]  M. Wu, M. Burke, S. Son, and R. Yetter, "Flame acceleration and the transition to detonation of stoichiometric ethylene/oxygen in microscale tubes," Proc. Combust. Inst. **31**, 2429 (2007).

[8]  G. Searby and P. Clavin, "Weakly turbulent, wrinkled flames in premixed gases," Combust. Sci. Technol. **46**, 167 (1986).

[9]  N. Peters, *Turbulent Combustion* (Cambridge University Press, New York, NY, 2000).

[10] S. Ishizuka, "Flame propagation along a vortex axis," Prog. En. Combust. Sci. **28**, 477 (2002).

[11] V. Akkerman, V. Bychkov, and L. E. Eriksson, "Numerical study of turbulent flame velocity," Combust. Flame **151**, 452 (2007).

[12] V. V. Bychkov and M. A. Liberman, "Dynamics and stability of premixed flames," Phys. Rep. **325**, 115 (2000).

[13] S. Kadowaki and T. Hasegawa, "Numerical simulation of dynamics of premixed flames: flame instability and vortex-flame interaction," Prog. En. Combust. Sci. **31**, 193 (2005).

[14] D. Bradley, T. M. Cresswell, and J. S. Puttock, "Flame acceleration due to flame induced instabilities in large-scale explosions," Combust. Flame **124**, 5551 (2001).

[15] G. Jomaas and C.K. Law, Proc. 46th AIAA Aerospace Sciences Meeting and Exhibit, 7-10 January 2008, Reno, Nevada, paper 1414.

[16] C. Clanet and G. Searby, "On the theory of tulip flame phenomena," Combust. Flame **105**, 225 (1996).

[17] V. Bychkov, V. Akkerman, G. Fru, L. E. Eriksson, and A. Petchenko, "Flame acceleration in the early stages of burning in tubes," Combust. Flame **150**, 263 (2007).

[18] G. Searby, "Acoustic instability in premixed flames," Combust. Sci. Technol. **81**, 221 (1992).

[19] V. Bychkov, "Analytical scalings for flame interaction with sound waves," Phys. Fluids **11**, 3168 (1999).

[20] L. Kagan and G. Sivashinsky, "The transition from deflagration to detonation in thin channels," Combust. Flame **134**, 389 (2003).

[21] J. D. Ott, E. S. Oran, and J. D. Anderson, "A mechanism for flame acceleration in narrow tubes," AIAA Journal **41**, 1391 (2003).

[22] V. Bychkov, A. Petchenko, V. Akkerman, and L. E. Eriksson, "Theory and modeling of accelerating flames in tubes," Phys. Rev. E **72**, 046307 (2005).

[23] V. Akkerman, V. Bychkov, A. Petchenko, and L. E. Eriksson, "Accelerating flames in cylindrical tubes with nonslip at the walls," Combust. Flame **145**, 206 (2006).

[24] D. Valiev, V. Bychkov, V. Akkerman, L. E. Eriksson, and M. Marklund, "Heating of the fuel mixture due to viscous stress ahead of accelerating flames in deflagration-to-detonation transition," Phys. Lett. A **372**, 4850 (2008).

[25] V. Gamezo, T. Ogawa, and E. S. Oran, "Flame acceleration and DDT in channels with obstacles: Effect of obstacle spacing," Combust. Flame **155**, 302 (2008).

[26] V. Bychkov, D. Valiev, and L. E. Eriksson, "Physical mechanism of ultrafast flame acceleration," Phys. Rev. Lett. **101**, 164501 (2008).

[27] V. Akkerman, V. Bychkov, A. Petchenko, and L. E. Eriksson, "Flame oscillations in tubes with nonslip at the walls," Combust. Flame **145**, 675 (2006).

[28] A. Petchenko, V. Bychkov, V. Akkerman, and L. E. Eriksson, "Violent folding of a flame front in a flame-acoustic resonance," Phys. Rev. Lett. **97**, 164501 (2006).

[29] A. Petchenko, V. Bychkov, V. Akkerman, and L. E. Eriksson, "Flame-sound interaction in tubes with nonslip walls," Combust. Flame **149**, 418 (2007).

[30] T. Poinsot and D. Veynante, *Theoretical and Numerical Combustion* ($2^{nd}$ ed., Edwards, Ann Arbor, MI, 2005).

[31] L. E. Eriksson, *Development and validation of highly modular solver versions G2DFlow and G3DFlow series for compressible viscous reactive flow* (VAC Report No. 9970-1162, Volvo Aero Corp., 1995).

[32] C. Wollblad, L. Davidson, and L. E. Eriksson, "Large eddy simulation of transonic flow with shock wave/turbulent boundary layer interaction," AIAA Journal **44**, 2340 (2006).

[33] N. Andersson, L. E. Eriksson, and L. Davidsson, "Large-eddy simulation of subsonic turbulent jets and their radiated sound,"AIAA Journal **43**, 1899 (2005).

[34] D. Valiev, V. Bychkov, V. Akkerman, and L. E. Eriksson, "Different stages of flame acceleration from slow burning to Chapman-Jouguet deflagration", Phys. Rev. E **80**, 036317 (2009).

[35] V. Bychkov, V. Akkerman, D. Valiev, and C. K. Law, "Role of compressibility in moderating flame acceleration in tubes", Phys. Rev. E **81**, 026309 (2010).

**FIGURE CAPTIONS**

**Fig. 1.** Accelerating or hypothetical steady flame in a tube/channel with non-slip walls and both ends open.

**Fig. 2.** Scaled profile of the flow velocity, Eq. (25), for various $\mu = 0.1, 3, 6, 15$.

**Fig. 3.** Scaled flame shape, Eq. (37), for $\Theta = 8$ and various $\text{Re} = 10, 50, 200$.

**Fig. 4.** Flame acceleration in a cylindrical tube of radius (a) $R = 10 L_f$ and (b) $R = 30 L_f$. Both ends of the tube are open. The flame isotherms are taken from $600\,\text{K}$ to $2100\,\text{K}$ with the step of $300\,\text{K}$ in each plot in both figures. (a) The positions (I) – (VIII) are related to the time instants $t\, S_L / R = 0 \sim 1.4$, with equal time intervals $\Delta t\, S_L / R = 0.2$. (b) The positions (I) – (VII) are related to the time instants $t\, S_L / R = 0 \sim 1.8$, with equal time intervals $\Delta t\, S_L / R = 0.3$.

**Fig. 5.** Flame oscillations in a 2D channel of width $D = 2R = 40 L_f$ with both ends open [27]. The flame isotherms are taken from $600\,\text{K}$ to $2100\,\text{K}$ with the step of $300\,\text{K}$ in each plot. The positions (a) - (f) are related to the time instants $t\, S_L / R = 0, 0.57, 1.13, 1.7, 2.26, 2.83$.

**Fig. 6.** Flame acceleration in a cylindrical tube of radius $R = 25 L_f$ with one end closed [23]. The flame isotherms are taken from $600\,\text{K}$ to $2100\,\text{K}$ with the step of $300\,\text{K}$ in each plot. The positions (a) – (g) are related to the time instants $t\, S_L / R = 0 \sim 0.72$, with interval $\Delta t\, S_L / R = 0.12$.

**Fig. 7.** Evolution of the flame shape in a cylindrical tube of radius $R = 30L_f$ with both ends open. The colors designate the temperature: from $300\,\text{K}$ in the cold gas to $2400\,\text{K}$ in the burnt matter. The snapshots (a) – (g) are related to the time instants $tS_L/R = 0.4 \sim 2.8$, with equal time intervals $\Delta t\, S_L/R = 0.4$.

**Fig. 8.** The scaled total flame velocity $U_w/S_L$ versus time for open cylindrical tubes with $\Theta = 8$ and $Pe = 5 \sim 35$. The plots are related to three main stages of the flame dynamics: (a) initial (concave), (b) intermediate (transitional) and (c, d) final (self-similar, convex). Figure 8d is a counterpart of Fig. 8c in the semi-logarithmic scale. The dashed plot is related to the test simulation run with $Pe = 10$ and the simulation grid $0.2L_f \times 0.2L_f$. Two dotted plots in Figs. 8 c, d are related to a tube with a closed end [23].

**Fig. 9.** Acceleration rate versus the Peclet number for $\Theta = 8$. The solid plot shows the numerical solution to Eq. (39). The dashed plot presents the zeroth-order approximation, Eq. (43). The simulation results are shown by symbols. The dotted plot ($\chi = 1$) is related to a tube with a closed end [23].

**Fig. 10.** The scaled total burning rate $U_w/S_L$ versus time for an open cylindrical tube with $Pe = 10$ and various expansion factors $\Theta = 2 \sim 8$.

**Fig. 11.** The scaled total burning rate $U_w/S_L$ versus time for an open cylindrical tube with $Pe = 20$ and various expansion factors $\Theta = 2 \sim 8$. The plots are related to: (a) concave and (b) convex stages of the flame dynamics.

**FIGURES**

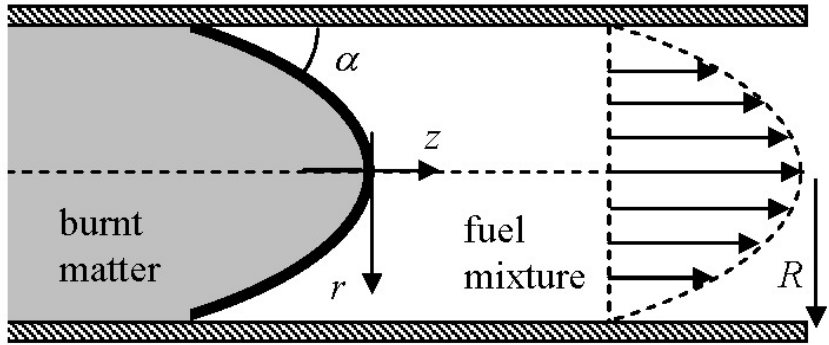

**Fig. 1.** Accelerating or hypothetical steady flame in a tube/channel with non-slip walls and both ends open.

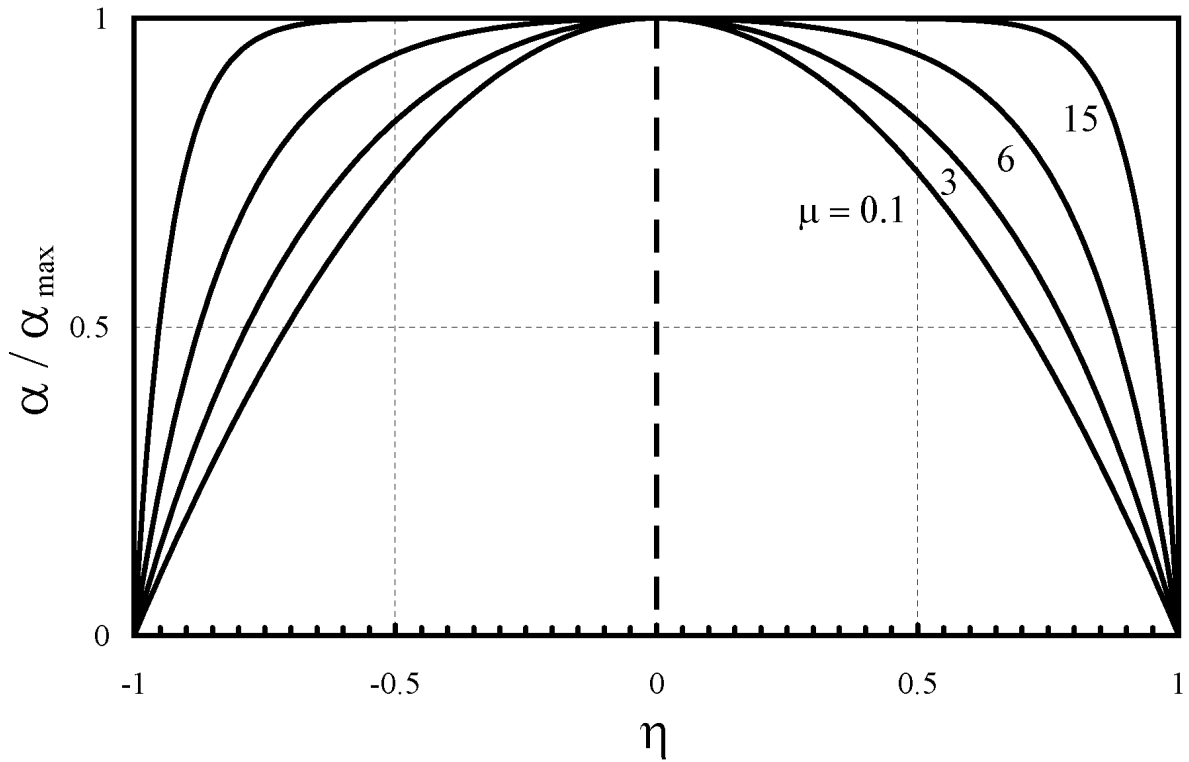

**Fig. 2.** Scaled profile of the flow velocity, Eq. (25), for various $\mu = 0.1, 3, 6, 15$.

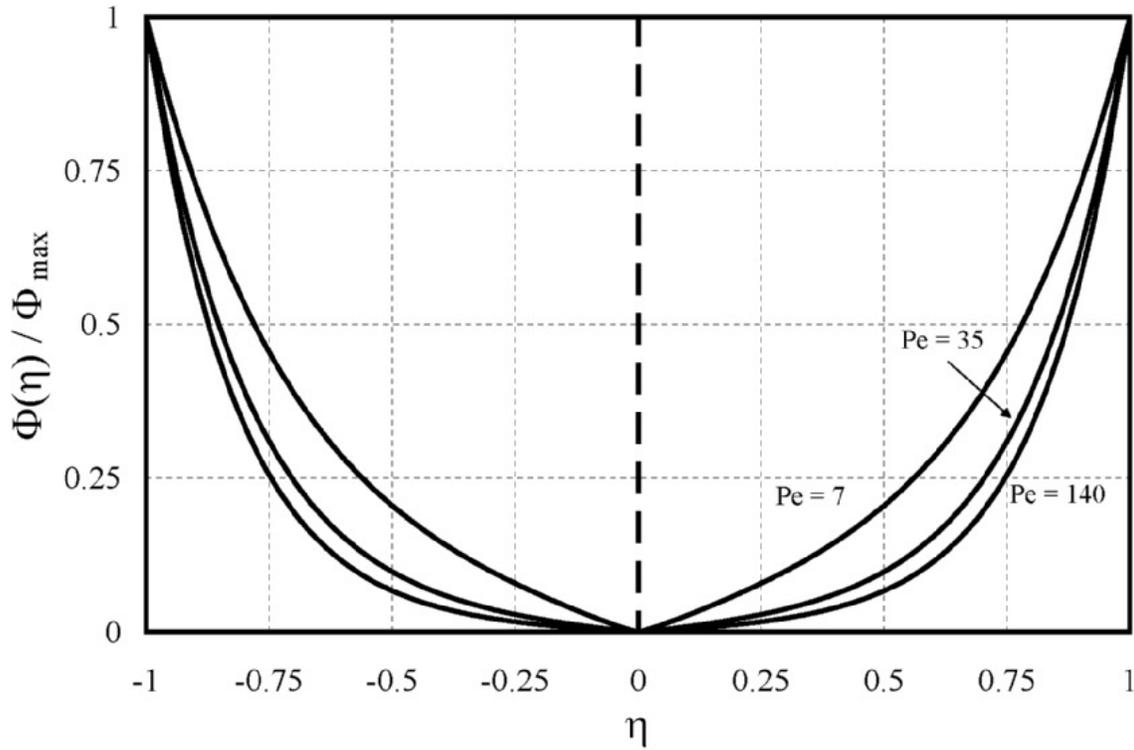

**Fig. 3.** Scaled flame shape, Eq. (37), for $\Theta = 8$ and various $Re = 10, 50, 200$.

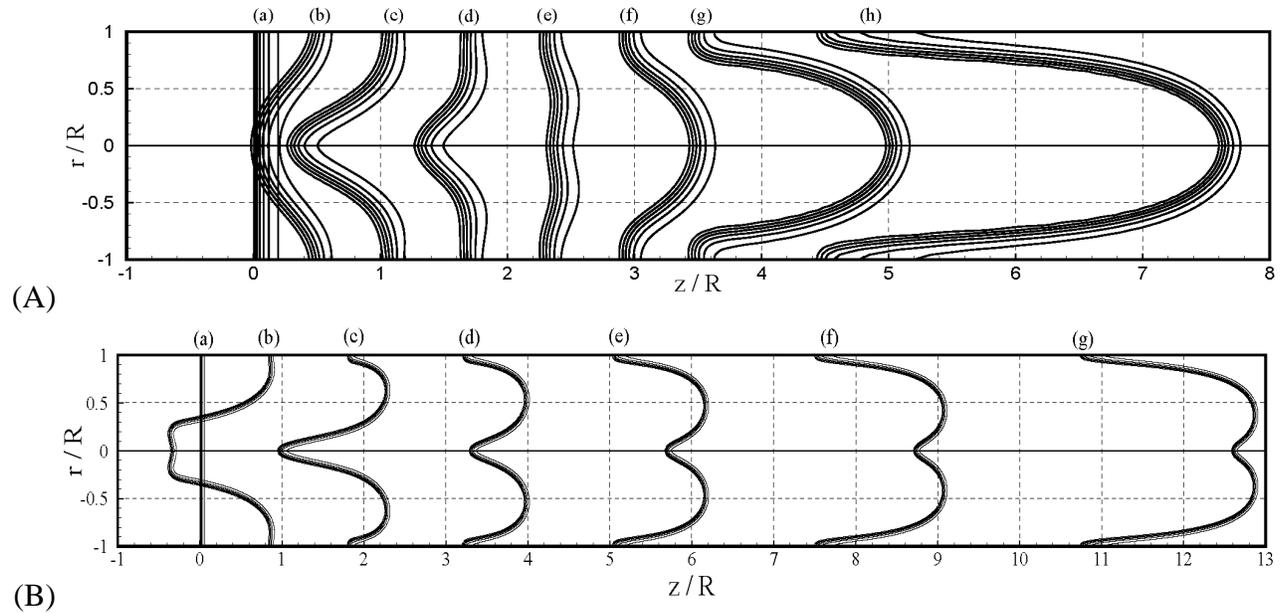

**Fig. 4.** Flame acceleration in a cylindrical tube of radius (a) $R = 10L_f$ and (b) $R = 30L_f$. Both ends of the tube are open. The flame isotherms are taken from $600\,\text{K}$ to $2100\,\text{K}$ with the step of $300\,\text{K}$ in each plot in both figures. (a) The positions (I) – (VIII) are related to the time instants $tS_L/R = 0 \sim 1.4$, with equal time intervals $\Delta t\, S_L/R = 0.2$. (b) The positions (I) – (VII) are related to the time instants $tS_L/R = 0 \sim 1.8$, with equal time intervals $\Delta t\, S_L/R = 0.3$.

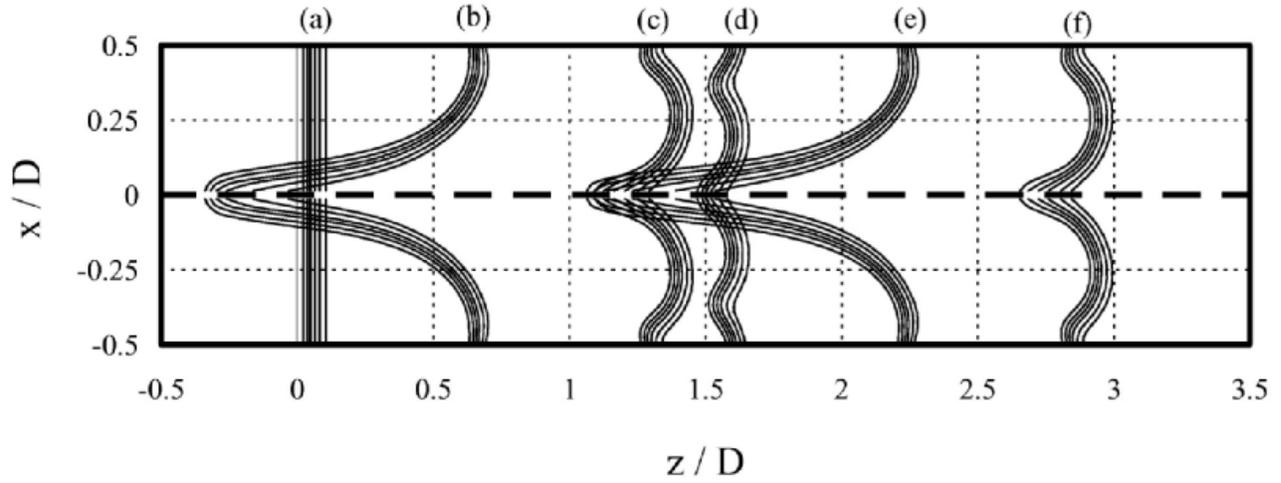

**Fig. 5.** Flame oscillations in a 2D channel of width $D = 2R = 40L_f$ with both ends open [27]. The flame isotherms are taken from $600\,\text{K}$ to $2100\,\text{K}$ with the step of $300\,\text{K}$ in each plot. The positions (a) - (f) are related to the time instants $t\,S_L/R = 0, 0.57, 1.13, 1.7, 2.26, 2.83$.

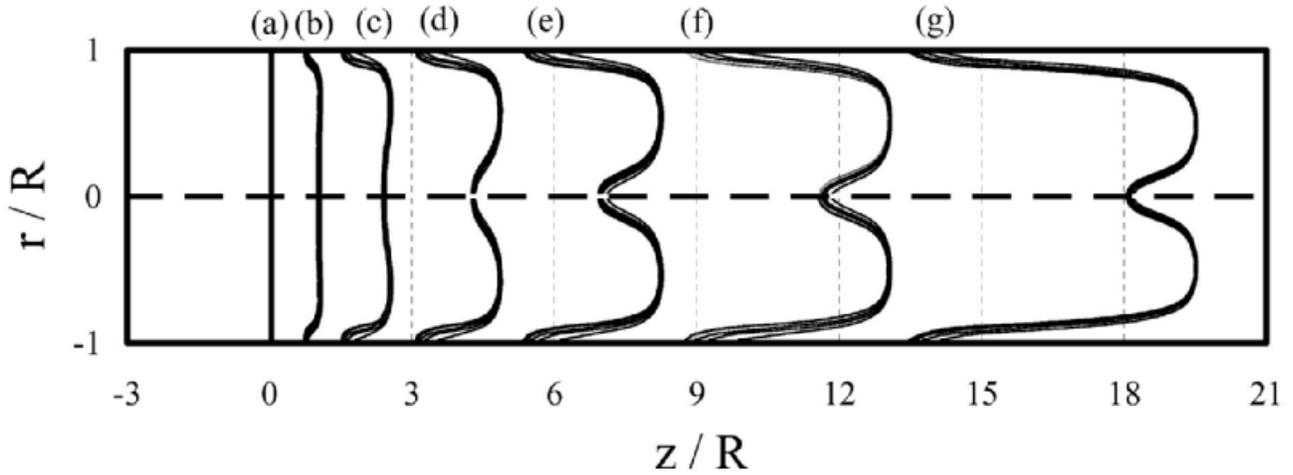

**Fig. 6.** Flame acceleration in a cylindrical tube of radius $R = 25L_f$ with one end closed [23]. The flame isotherms are taken from $600\,\text{K}$ to $2100\,\text{K}$ with the step of $300\,\text{K}$ in each plot. The positions (a) – (g) are related to the time instants $t\,S_L/R = 0 \sim 0.72$, with interval $\Delta t\,S_L/R = 0.12$.

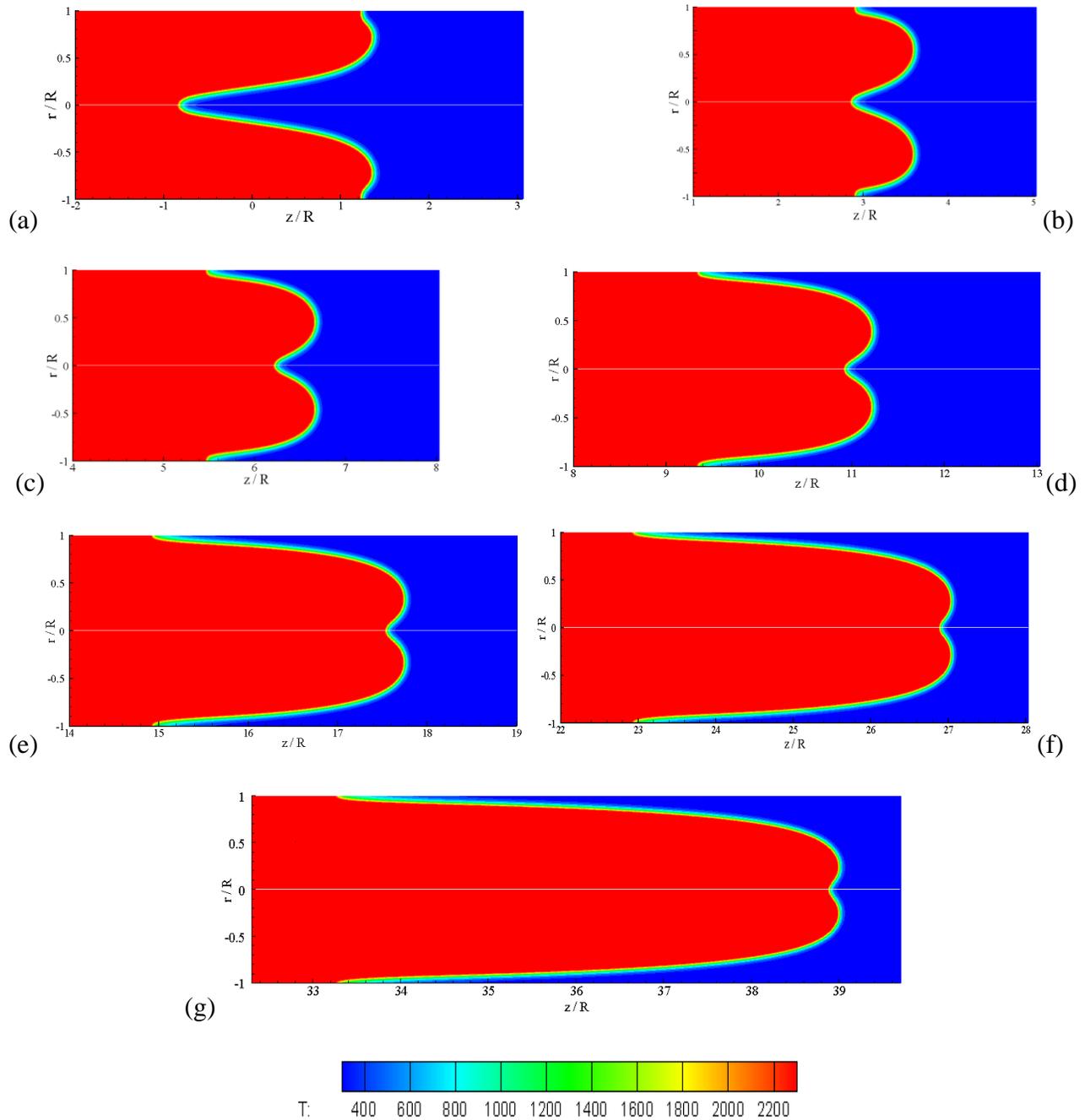

**Fig. 7.** Evolution of the flame shape in a cylindrical tube of radius $R = 30L_f$ with both ends open. The colors designate the temperature: from $300\,\text{K}$ in the cold gas to $2400\,\text{K}$ in the burnt matter. The snapshots (a) – (g) are related to the time instants $tS_L/R = 0.4 \sim 2.8$, with equal time intervals $\Delta t\, S_L / R = 0.4$.

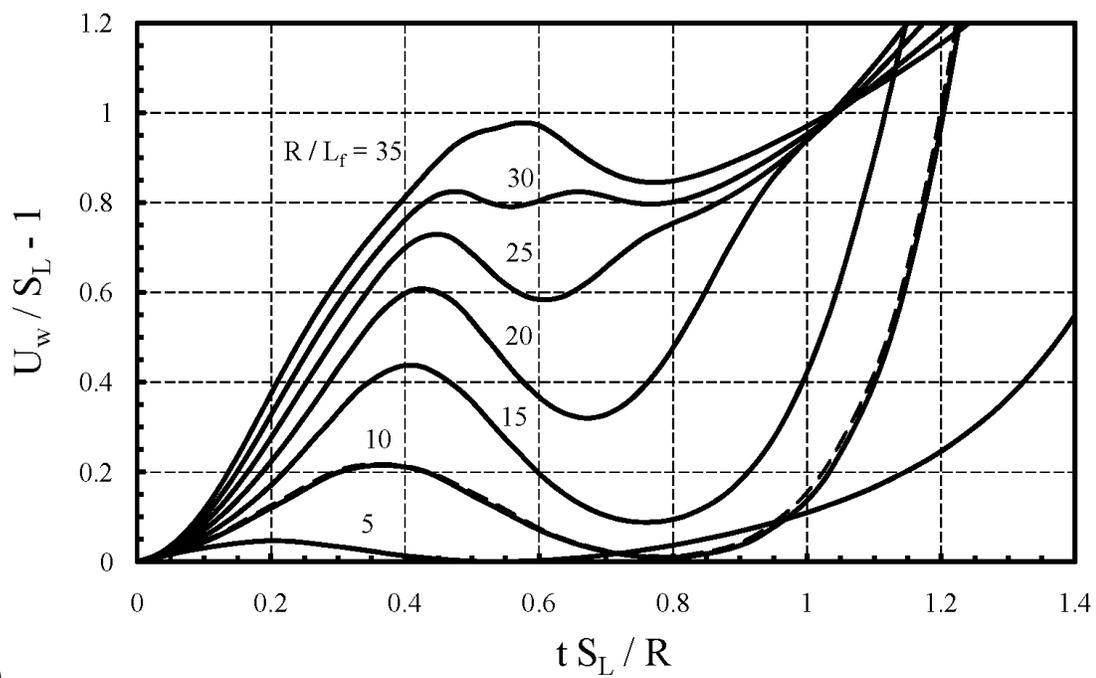

(A)

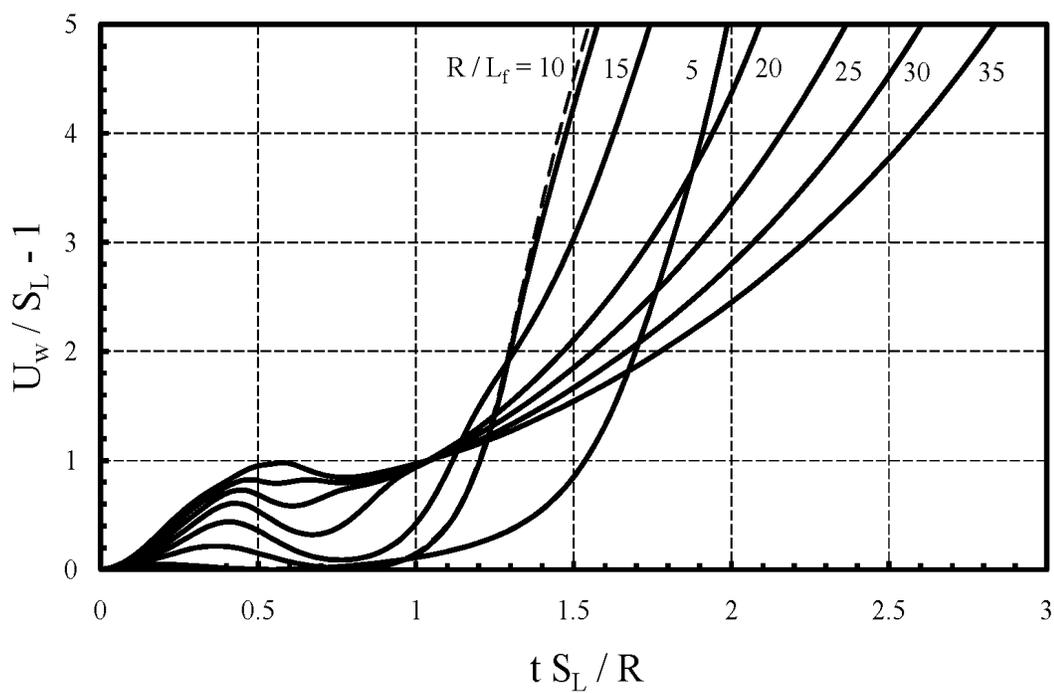

(B)

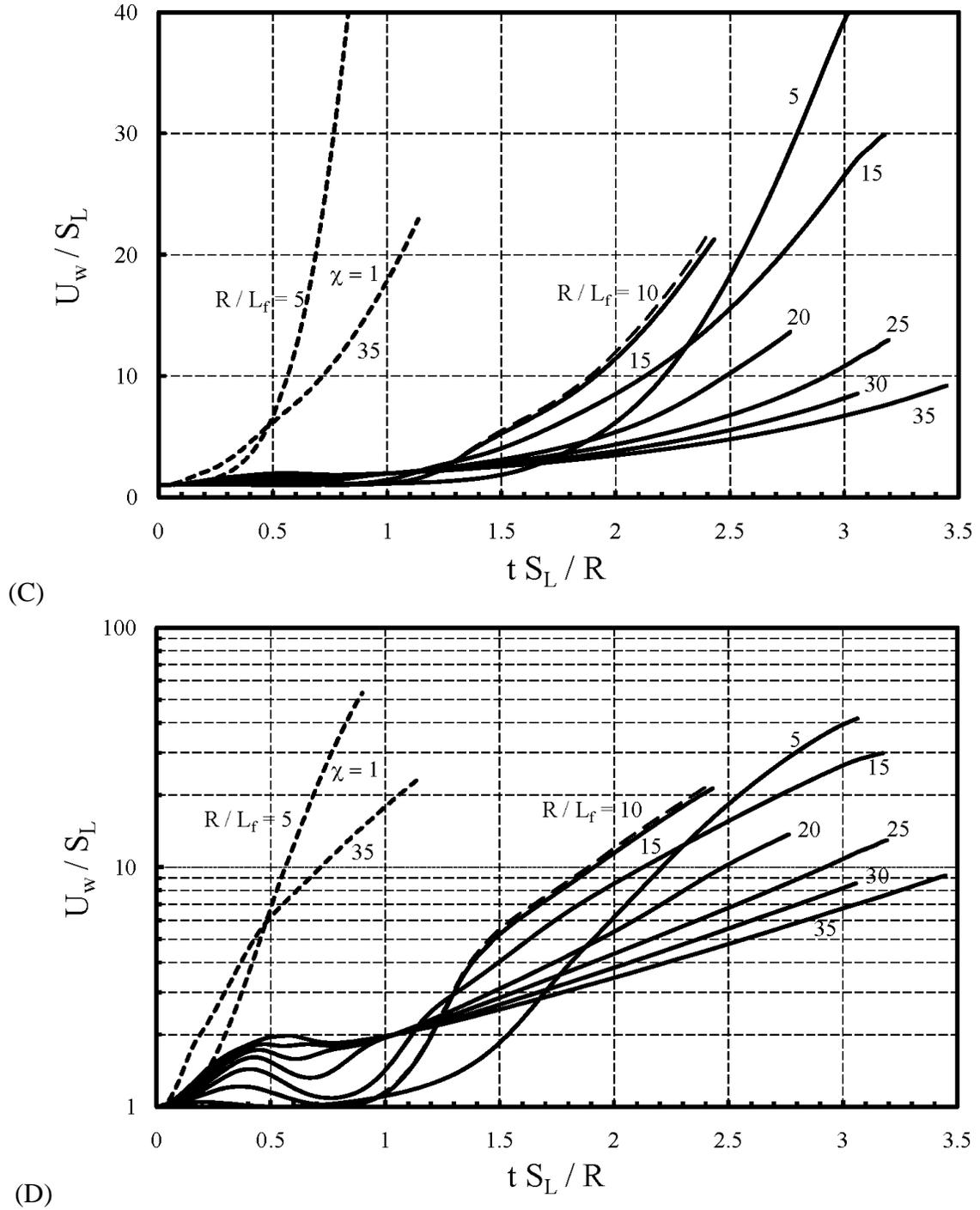

**Fig. 8.** The scaled total flame velocity $U_w / S_L$ versus time for open cylindrical tubes with $\Theta = 8$ and $Pe = 5 \sim 35$. The plots are related to three main stages of the flame dynamics: (a) initial (concave), (b) intermediate (transitional) and (c, d) final (self-similar, convex). Figure 8d is a counterpart of Fig. 8c in the semi-logarithmic scale. The dashed plot is related to the test simulation run with $Pe = 10$ and the simulation grid $0.2L_f \times 0.2L_f$. Two dotted plots in Figs. 8 c, d are related to a tube with a closed end [23].

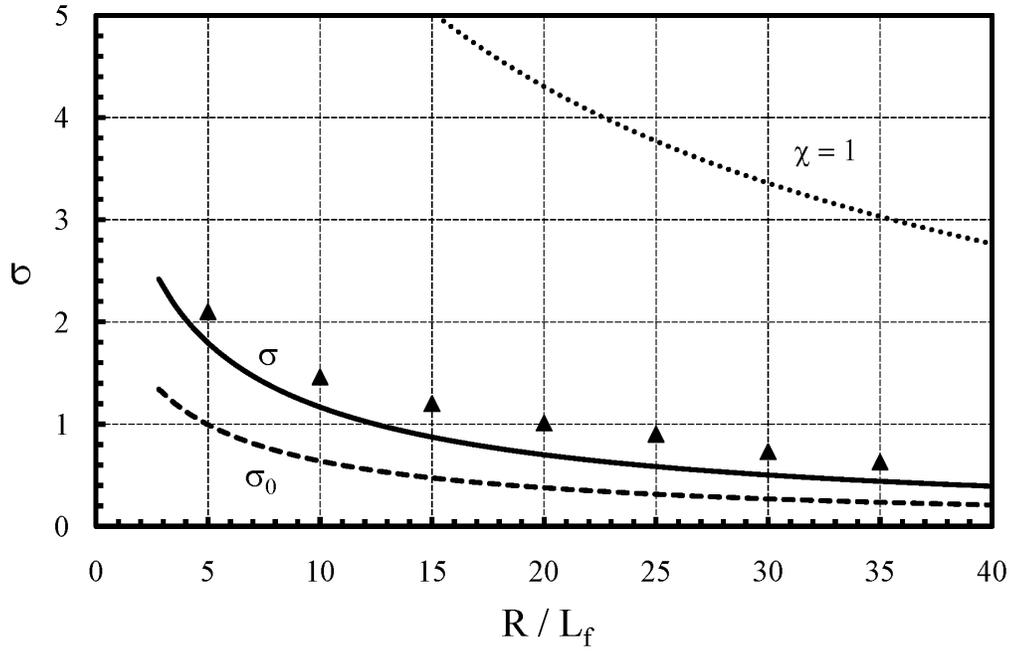

**Fig. 9.** Acceleration rate versus the Peclet number for $\Theta = 8$. The solid plot shows the numerical solution to Eq. (39). The dashed plot presents the zeroth-order approximation, Eq. (43). The simulation results are shown by symbols. The dotted plot ($\chi = 1$) is related to a tube with a closed end [23].

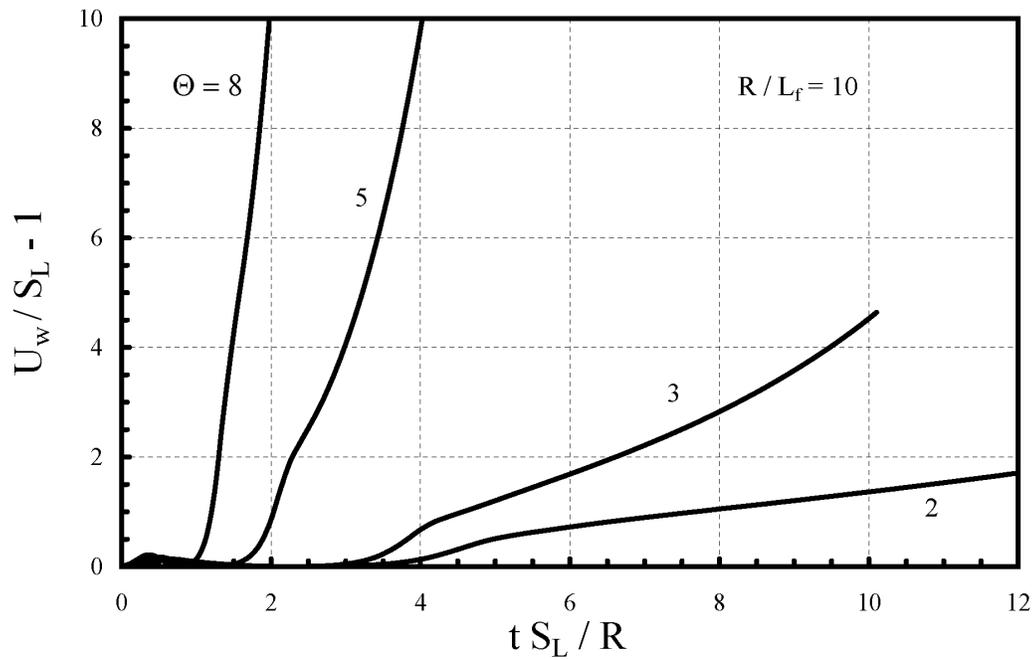

**Fig. 10.** The scaled total burning rate $U_w / S_L$ versus time for an open cylindrical tube with $Pe = 10$ and various expansion factors $\Theta = 2 \sim 8$.

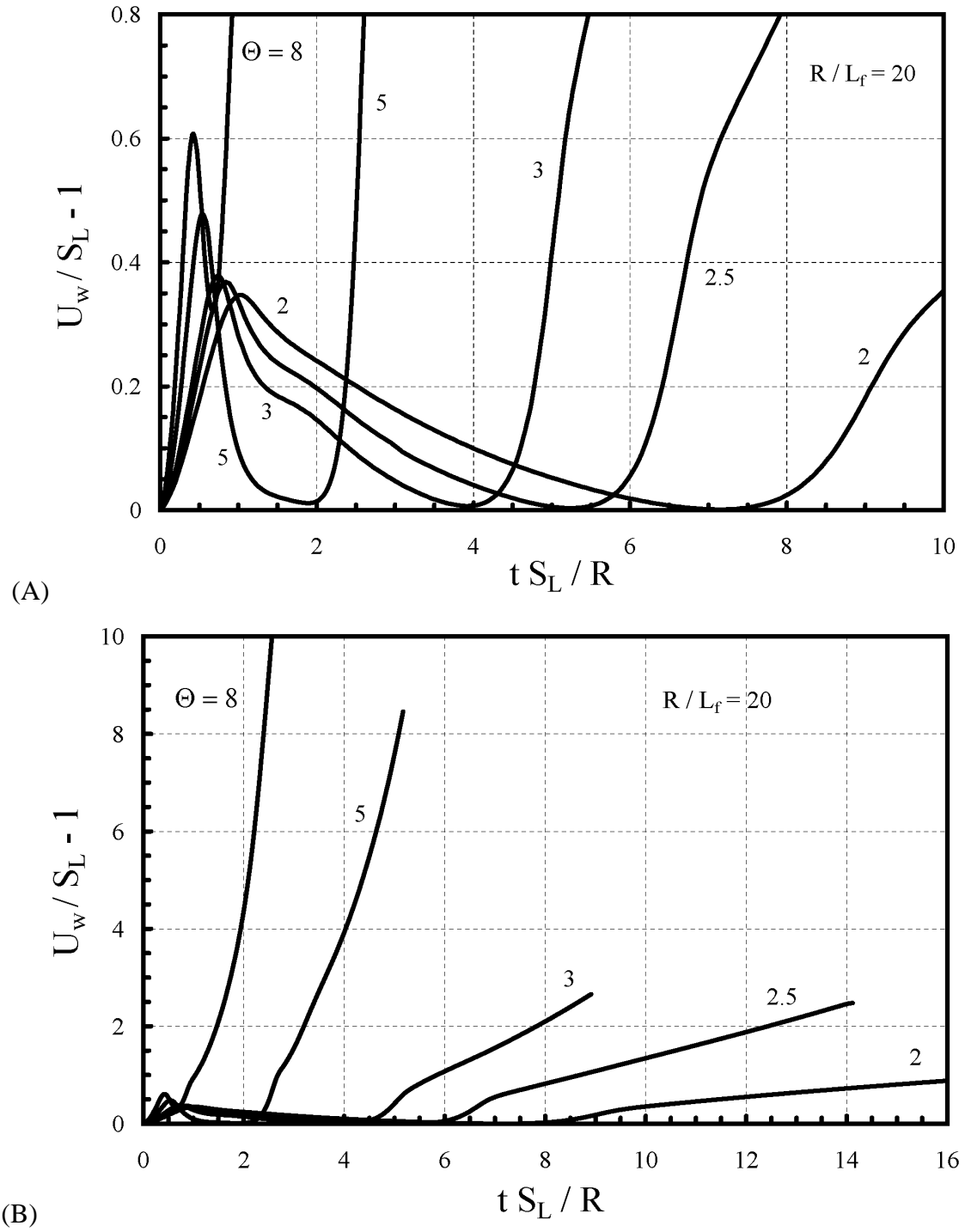

**Fig. 11.** The scaled total burning rate $U_w / S_L$ versus time for an open cylindrical tube with $Pe = 20$ and various expansion factors $\Theta = 2 \sim 8$. The plots are related to: (a) concave and (b) convex stages of the flame dynamics.